\documentclass{aa}
\usepackage{graphics,latexsym,amssymb,times,psfig}

\def\gsim{ \lower .75ex \hbox{$\sim$} \llap{\raise .27ex \hbox{$>$}} } 
\def\lsim{ \lower .75ex\hbox{$\sim$} \llap{\raise .27ex \hbox{$<$}} } 

\newcommand{\pim}{$\pm$}

\begin{document}

\title{The clustering of the luminosities of optical afterglows of long
Gamma Ray Bursts}


\author{M. Nardini \inst{1,2}, 
G. Ghisellini \inst{1},
G. Ghirlanda \inst{1}, 
F. Tavecchio \inst{1}, 
C. Firmani \inst{1,3},
D. Lazzati \inst{4}}
\offprints{G. Ghisellini; gabriele@merate.mi.astro.it}
\institute{
Osservatorio Astronomico di Brera, via Bianchi 46, I--23807 Merate, Italy.
\and
Univ. di Milano--Bicocca, P.za della Scienza 3, I--20126, Milano, Italy.
\and
Instituto de Astronom\'{\i}a, U.N.A.M., A.P. 70-264, 04510, M\'exico, D.F., M\'exico
\and
JILA, University of Colorado, Boulder, CO 80309-0440, USA
}

\date{Received 2005}
 
\titlerunning{Clustering of optical afterglow luminosities}
\authorrunning{M. Nardini et al.}

\abstract{
We studied the optical afterglows of the 24 pre--SWIFT
Gamma--Ray Bursts with known spectroscopic redshift and published 
estimates of the optical extinction in the source frame.
We find an unexpected clustering of the optical afterglow luminosities
measured 12 hours (source frame time) after the trigger.
For 21 out of 24 bursts, 
the distribution of the optical luminosities is narrower than
the distribution of the X--ray luminosities, and even narrower than
the distribution of the
ratio between the monochromatic optical luminosities
and the total isotropic emitted prompt energy.
Three bursts stand apart from the distribution of the other
sources, being underluminous by a factor $\sim$ 15.
We compare this result with the somewhat analogous result concerning
the luminosity of the X--ray afterglows studied by Gendre \& Bo\"er.
For all our GRBs we construct the optical to X--ray spectral energy 
distribution. 
For all but a minority of them, the optical and the X--ray emissions 
are consistent with being produced by the same radiation process.
We discuss our results in the framework of the ``standard" external 
shock synchrotron model.
Finally, we consider the behavior of the first GRBs of known 
redshifts detected by SWIFT.
We find that these SWIFT GRBs entirely confirm our findings.
}
\maketitle

\section{Introduction}

The most common approach to directly compare the afterglow emission
of different bursts, is to compute the light curves in the observer
reference frame, in terms of their {\it fluxes} vs the
{\it observed} time $t_{\rm obs}$.
However, when the redshift is known, a more fruitful approach
is to compare the light curves of the {\it luminosities} 
of different bursts, using the rest frame time 
$t_{\rm RF}=t_{\rm obs}/(1+z)$.
Although such attempts have already been done in the past
(see, e.g. Gendre \& Bo\"er 2005, hereafter GB05; Kumar \& Piran 2000)
they concerned mainly the X--ray luminosities of relatively small 
samples of GRBs, and not the optical luminosities 
 
(but see Berger et al. 2005 for a study of the first SWIFT bursts).
 
From these earlier studies (see also Piran et al. 2000) it appeared 
that the X--ray afterglow luminosities (calculated at the same 
time in the rest frame) were characterized by a smaller dispersion 
than the dispersion of the total energies radiated during the 
prompt emission (but see Berger, Kulkarni \& Frail 2003 for 
a different conclusion).
In addition, Bo\"er \& Gendre (2000) and GB05 found that the 
X--ray afterglow luminosities showed the tendency to cluster 
into two groups (different by a factor $\sim$30 in luminosity)
with a  small dispersion within each group.

These authors tried also to draw conclusions from the
optical luminosity but were not successful because the absorption 
was largely unknown at that time, and because of a too small sample.

These earlier results prompted us to study the behavior
of the optical afterglow luminosities.
One of the main initial motivations of our study was the possibility
that what it seems to be a ``dichotomy" in the X--ray afterglow luminosity
could be present also in the optical, therefore helping to understand
the problem of the so called ``dark" bursts (bursts with a detected
X--ray afterglow but no optical detection).
Consider also that De Pasquale et al. (2003), comparing
GRBs (all with detected X--ray afterglow) with and without optical
detection, found that ``dark" GRBs tend to be fainter in the X--rays,
by a factor $\sim 5$ in flux at the same observed time.
We however expect a dispersion of the optical luminosity 
(at a given rest frame time) greater than the corresponding 
dispersion of the X--ray luminosities.
Electrons emitting X--rays by the synchrotron process, in fact, likely 
cool in a dynamical timescale (also several hours after trigger), 
and this  implies (in the standard synchrotron fireball model) 
that the emitted X--ray emission is insensitive to the density of the 
external medium producing the external shocks.
On the contrary, it is likely that electrons emitting in the optical 
do not cool in a dynamical timescale (after about a day since trigger), 
and therefore the optical emission does depend on the density of 
the circumburst material.
If the dispersion introduced by this effect is not too large,
some sort of ``dichotomy" could survive, and then could flag
the existence of two families of GBRs with two different 
average afterglow luminosities.
Dark GRBs could then be thought to belong to the underluminous family,
therefore more difficult to detect, and more so in the optical, 
if some extinction in the host galaxy is present.


The results presented in the following are instead 
quite puzzling, since, contrary to the simple expectations
mentioned above, the optical afterglow luminosities show
a degree of clustering which is tighter
than that shown by the X--ray afterglow luminosities.
We indeed find an indication (albeit still weak, due to the
small statistics) of a dichotomy in the optical 
luminosity distributions.
But, more intriguingly, we find an unexpected optical luminosity 
clustering of the 
large majority of the bursts analyzed by us (21 out of 24).
In order to understand it, we have  constructed, 
for all GRBs of our sample, the optical to X--rays Spectral 
Energy Distribution (SED) at a given time, assembling
spectral information contained in the multiband photometry 
in the optical and the X--ray continuum spectra.
This allows to see if  both the optical and the
X--ray fluxes are consistent with being produced by
a single electron population by the synchrotron process,
or if there is some indications of X--ray fluxes being
produced by an additional component (i.e. a possible emergent 
inverse Compton flux in the X--ray band), being possibly 
responsible for the larger dispersion of X--ray luminosities 
with respect to the optical ones.
As we will show, this is not the case for most of the sources.

We also consider 
 
in Section 5

the burst detected by SWIFT and for which
the redshift is known.
All of them but GRB 050401, GRB 050525 and GRB 050730 
lack information about the optical absorption  in their hosts. 
With this caveat, we calculate their optical luminosities 
and find that they are consistent with the clustering
properties of the other bursts.
Instead, we find that, on average, they are more
powerful in X--rays with respect to the pre--SWIFT bursts
and therefore they broaden the X--ray luminosity distribution.

We finally discuss these results in the framework of the 
standard fireball external shocks synchrotron scenario,
and the possible implications for dark bursts.

Throughout this paper, we adopt a cosmology with $\Omega_{\rm M}=0.3$ 
and $\Omega_{\Lambda}=h_0=0.7$.

\begin{table*}
\begin{center}
\begin{tabular}{l l c c c c c c c}
\hline\hline\noalign{\smallskip}
GRB  & $z$& ref z  & $\beta$ & $A_R^{\rm Gal}$ & $A_V^{\rm host}$ &
$A_{R(1+z)}^{\rm host}$ 
& $\log L^{\rm 12h}_{\nu_R}$ & Ref\\
\hline
\object{970508} & 0.835 &me97 &  1 & 0.13  & 0 & 0 & 30.42  & ga98 \\ 
\object{971214} & 3.418 &ku98& 1.03$\pm$0.18 & 0.04   & 0.38$\pm$0.08 & 0.99 & 30.39 & wi99\\
\object{980613} & 1.0964 &dj99& 0.59$\pm$0.03 & 0.23   &  & 0.45 &29.31 & hj02  \\
\object{980703} & 0.966  &dj98& 1.01\pim0.01  & 0.15  & 1.51$\pm$0.11 & 2.50  & 30.82& vr99 \\  
                &     &   & 0.78          & 0.15  & 0.90$\pm$0.20 & 1.48 & 30.34 & bl98 \\
\object{990123} & 1.6 &ke99& 0.750\pim0.068 & 0.04 & 0 & 0 & 30.62 & ho00 \\
\object{990510} & 1.619 &vr99b& 0.49$\pm$0.1 & 0.54 & 0  & 0    & 30.73 &ho00 \\
                &   &    & 0.55\pim0.1  & 0.48 &    &      & 30.75 & be99\\  
\object{991216} & 1.02 &vr99c& 0.58$\pm$0.08 & 1.67 & 0 & 0 & 30.89 & ga00 \\
\object{000301c} & 2.0670 &fe00& 0.70$\pm$0.09 & 0.13 & 0.09$\pm$0.04 & 0.26 & 30.99 & je01\\ 
\object{000418} & 1.1181  &bl00& 0.81& 0.08 & 0.96$\pm$0.20 & 1.69 & 30.71 & kl00 \\
\object{000911} & 1.06 &pr02 & 1.3 & 0.30 & 0.39 & 0.69 & 30.66  & ma05 \\
\object{000926} & 2.0375 &fy00& 1.0$\pm$0.2 &  0.06 & 0.18$\pm$0.06 & 0.53 &31.06 & fy01  \\
\object{010222} & 1.477 &jh01b& 1.07$\pm$0.09 & 0.06 & 0    & 0    & 30.52  & st01 \\
                &       & &0.89\pim0.03  &      & 0    & 0    & 30.45  & jh01\\
                &       & &0.5           &      & 0.19 & 0.42 & 30.46  & le01\\
\object{010921} & 0.45 &dj01& p=3.03 & 0.396 & 1.16$\pm$0.07 & 1.43 & 30.77 & pr02a\\
\object{011121} & 0.36 & in01&0.62\pim0.05 &1.32  & 0 & 0 & 29.65 & gr03 \\
                &      && 0.76\pim0.15 &0.97  & 0 & 0 & 29.58 & pr02b  \\
                &      && 0.66\pim0.13 &1.12  & 0 & 0 & 29.53 & ga03 \\
\object{011211} & 2.14 &gl01& 0.56\pim0.19 & 0.11 & 0.08$\pm$0.08 & 0.23 & 30.36 & ja03\\
                &      & &0.61\pim0.15 &      & 0.06          & 0.177& 30.29 & ho02\\
\object{020124} & 3.198 &at05& 1.32\pim0.25 & 0.14 & 0            & 0    & 30.22 & hj03 \\
                &       && 0.31\pim0.43 &      & 2.66\pim0.16 & 0.73 & 29.81 & hj03\\
                &       && 0.91\pim0.14 &      & 0            & 0     & 30.00 &hj03\\ 
\object{020405} & 0.69 &ma02& 1.45 & 0.15 & 0 & 0 & 30.44 & be03 \\
\object{020813} & 1.25 &fi02& 0.85$\pm$0.07 & 0.30 & 0.12\pim004 & 0.226 & 30.57 & co03 \\
\object{021004} & 2.3351 &gi02& 0.60\pim0.02 & 0.16 & 0.15 & 0.39 & 31.17 & pa03  \\  
\object{021211} & 1.004 &vr03& 0.55$\pm$0.10 & 0.07  &    &0.48 & 29.41  & fo03 \\
                &       & &0.69\pim0.14  &       & 0  & 0   & 29.27  & ho04\\ 
\object{030226} & 1.986 &gr03b& 0.55          & 0.05 & 0.52 & 0.98 & 30.59  &pa04\\ 
                &        & &0.70$\pm$0.03 &      & 0    & 0    & 30.27  & kl04 \\
\object{030323} & 3.3718 &vr03b& 0.89$\pm$0.04 & 0.13 & 0 & 0 & 30.92 & vr04 \\
\object{030329} & 0.1685 &ca03& 0.5        & 0.07 & 0.30$\pm$0.03 & 0.29 & 30.45 & bl04\\
                &        && 0.8\pim0.2 &       & 0.12         & 0.12 & 30.40 & ma03\\
\object{030429} & 2.66   & we03 & 0.36$\pm$0.12 & 0.165 & 0.34$\pm$0.04 & 0.99 & 30.78 & ja04\\
\hline 
\end{tabular}
\caption{ Sample of GRBs with measured redshift $z$ and estimated
host extinction $A_{V}^{\rm host}$. The optical spectral index
$\beta_{o}$ and the Galactic R--band extinction $A_{R}^{\rm Gal}$ are
reported. $A_{R(1+z)}^{\rm host}$ represents the host rest frame
R--band extinction and $\log L^{\rm 12h}_{\nu_R}$ is the rest frame
R--band luminosity calculated at 12 h (rest frame) according to Eq.~1.
References are given for $\beta_{o}$,$A_{R}^{\rm Gal}$, $A_{V}^{\rm
host}$ and $A_{R(1+z)}^{\rm host}$.~ References: me97: Metzger et al.,
1997; ga98: Garcia et al., 1998; ku98: Kulkarni et al., 1998; wi99:           
Wijers et al. 1999; dj99: Djorgovski et al., 1999; hj02: Hjorth et
al., 2002; dj98: Djorgovski et al., 1998; bl98: Bloom et al., 1998; 
ke99: Kelson et al., 1999; vr99: Vreeswijck et al. 1999; ho00: Holland
et al., 2000; vr99b: Vreeswijck et al., 1999(GCN324);
be99: Beuermann et al., 1999; vr99c: Vreeswijck et al., 1999(GCN496);
ga00: Garnavich et al., 2000; bl00: Bloom et al., 2000;
fe00: Feng et al., 2000;
je01: Jensen  et al., 2001; 
kl00: Klose  et al., 2000; pr02: Price et al., 2002;  ma05:
Masetti et al., 2005; fy00: Fynbo  et al., 2000; fy01: Fynbo  et al.,
2001;jh01b: Jha et al., 2001(GCN974); st01: Stanek et al., 
2001; jh01: Jha et al., 2001; le01: Lee et al.,
2001; pr02a: Price et al., 2002; dj01: Djorgovski et al., 2001;
gr03: Greiner et al., 2003;
pr02b: Price  et al., 2002; in01: Infante et al., 2001;
 ga03: Garnavich  et al., 2003; gl01: Gladders et al., 2001;
ja03: Jakobsson et al., 2003; ho02: Holland  et al., 2002; at05:
Atteia et al., 2005;
hj03: Hjorth et al., 2003; ma02: Masetti et al., 2002;  
be03: Bersier et al., 2003; co03: fi02: Iore et al., 2002;
Covino et al., 2003; gi02: Giannini et al., 2002;
pa03: Pandey  et al., 2003; vr03: Vreeswijck et al., 2003(GCN1785);
fo03: Fox et al.,
2003; ho04: Holland et al., 2004; vr03b:  Vreeswijck et al., 2003(GCN1953)
gr03b: Greiner et al., 2003(GCN1886);
vr04: Vreeswijk et al., 2004; ca03: Caldwell et al, 2003;
pa04: Pandey et al., 2004; kl04: Klose et al., 2004; bl04: Bloom et al.,
2004; ma03: Matheson et al., 2003; we03: Weidinger et al., 2003;
ja04: Jakobsson et al., 2004 }
\label{tabottico}
\end{center} 
\end{table*}
\begin{figure}
\vskip -0.5 true cm
\centerline{\psfig{figure=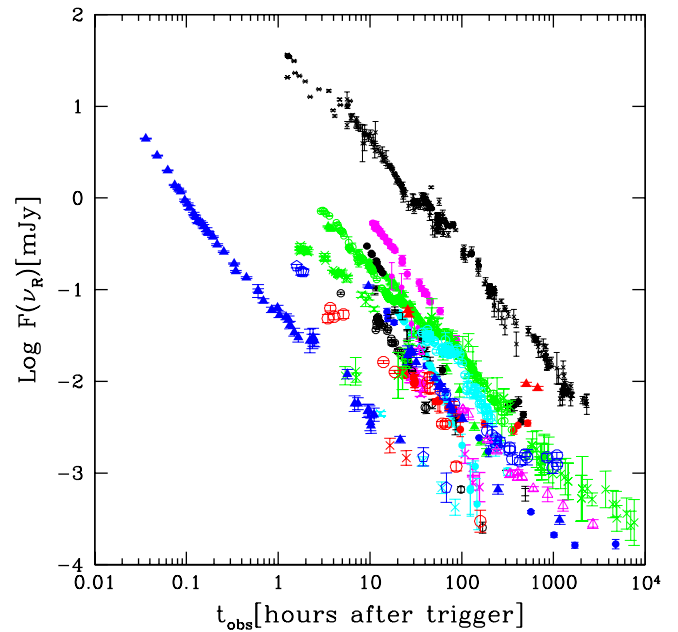,angle=0,width=10cm}}
\vskip -0.8 true cm
\caption{Light curves in terms of observed fluxes
versus observed time since the burst trigger for the 24 GRBs reported in
Tab. \ref{tabottico}. 
Fluxes have been corrected only for the galactic extinction.
The references for all the plotted data are given in the Appendix.
}
\label{f_obs}
\end{figure}
\begin{figure}
\vskip -0.5 true cm
\centerline{\psfig{figure=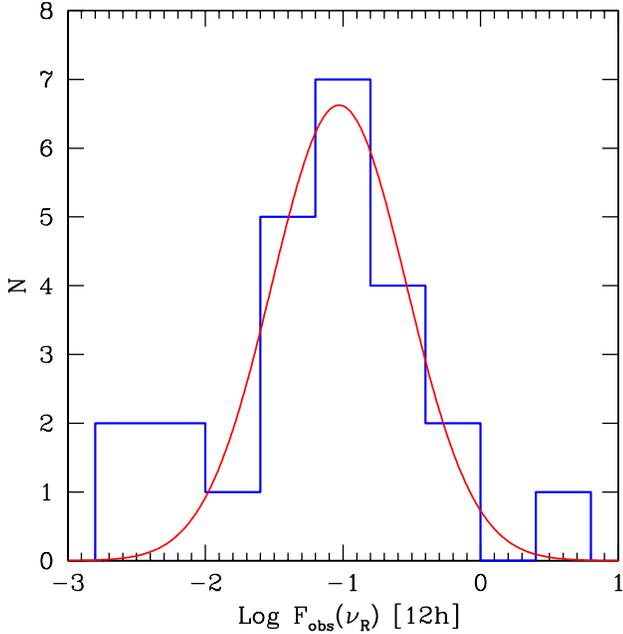,angle=0,width=10cm}}
\vskip -0.8 true cm
\caption{
Histogram of the observed fluxes (in mJy) in the $R$--band (Cousin system)
  12 hours after the trigger (observer frame -- from Fig.~\ref{f_obs}). All
  fluxes have been corrected for the foreground galactic extinction
  only. Superimposed to the histogram is a gaussian fit to the data
  with mean value $\mu=-1.03$ and dispersion $\sigma=0.48$.  
  Note that the fit is rather poor and most likely underestimates the real
  width of the distribution.
}
\label{istofobs}
\end{figure}

\begin{figure}
\vskip -0.5 true cm
\centerline{\psfig{figure=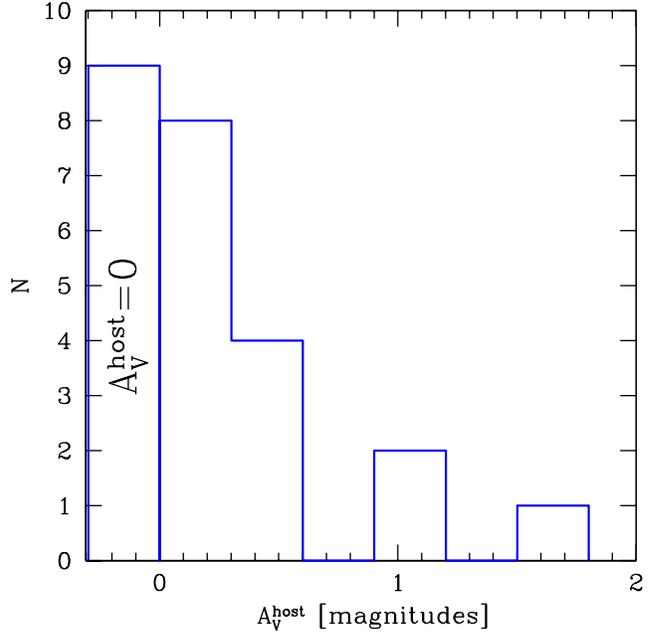,angle=0,width=10cm}}
\vskip -0.8 true cm
\caption{Histogram of the host absorption values $A_V^{\rm host}$ for
  the 24 GRBs of Tab. \ref{tabottico}. The 9 GRBs in the first bin have an
  optical spectrum which is consistent with a null host galaxy
  absorption.}
\label{A_vhost}
\end{figure}

\section{The sample}

To the aim of comparing the rest frame optical luminosities of
different GRBs, we applied all the relevant cosmological and
extinction corrections to the GRB light curves. 
In particular, one of our selection criteria is that the
absorption $A^{\rm host}_V$ in the host galaxy 

is known from the literature.


We have collected from the literature all GRBs of known spectroscopic
redshift $z$, detected optical afterglow, known optical spectral
index and known optical
extinction in the host rest frame $A^{\rm host}_V$. 
The total number of GRBs with measured redshifts is more than 50 
(as of July 31st, 2005) and 24 of those fulfill our selection criteria.
They are listed in Tab. \ref{tabottico}.
This list includes 13 out of the 17 GRBs present in the list of GB05. 
The 4 missing GRBs are: GRB 970228 (for which there is no
estimate of  $A^{\rm host}_V$), GRB 000210 and GRB 000214 (with no
detected optical afterglow), and GRB 980425 (an anomalous GRB associated
with the 1998bw supernova).

In Tab. \ref{tabottico} we report for every GRB of our sample the
redshift, the optical spectral index $\beta_o$, the galactic absorption 
$A^{\rm Gal}_V$ (taken from Schlegel (1998) 
maps except for GRB 011121 for which we report also
the values quoted in the corresponding references),
the host rest frame absorption $A^{\rm host}_V$, the absorption
$A_{\nu_R(1+z)}^{\rm host}$ at the rest frame frequency $\nu_R(1+z)$, 
the extinction and k--corrected monochromatic luminosity 
$L_{\nu_R}$ (at the source frame frequency corresponding to 
the $R$ band) and the references of the optical spectral 
index and extinction value.  

Fig. \ref{f_obs} shows the behavior of the observed
$R$--band fluxes as a function of the observed time $t_{\rm obs}$ for
all bursts listed in Tab. \ref{tabottico}. 
In this figure the fluxes are corrected for the galactic extinction only. 
In Fig. \ref{istofobs} we show the distribution of the observed fluxes
at the same observed time (12 hours after trigger).
Fitting the distribution of the observed optical fluxes 
with a gaussian
gives a (logarithmic) dispersion of $\sigma=0.48$ (see Tab. \ref{sigma}). 
Note that the gaussian fit is poor, and the real distribution could have
an even larger dispersion.

The monochromatic optical luminosities can be calculated  from
the observed monochromatic flux $F(\nu, t)$, by applying the
cosmological spectral and time corrections, as:
\begin{equation}
L(\nu, t )\, =\, 
{4\pi d_{\rm L}^2 F(\nu ,t) \over (1+z)^{1-\beta+\alpha}} 
\end{equation} 
where $d_{\rm L}$ is the luminosity distance and we assumed  
$F(\nu, t) \propto \nu^{-\beta}t^{-\alpha}$.  
Due to the much denser sampling in the Cousin $R$ band, we have 
assumed this band as the optical reference band for all the 
light curves
\footnote{
We appropriately convert Johnson $R$ magnitudes ($\lambda$=6800 \AA) 
into Cousin magnitudes when data in the former filter are given.}.  
We have then calculated all monochromatic
optical luminosities at the rest frame wavelength of 6400 \AA
~(corresponding to the Cousin $R$ filter)
\footnote{Note that Eq. (1) is
equivalent to Eq. 8 of Lamb \& Reichart (2000), who used the comoving
distance instead of the luminosity distance used here, and decay and
spectral index defined with an opposite sign.}.

All luminosities are given at the same rest frame time after trigger,
which we choose to be 12 hours.
This choice satisfies the following requirements: 
i) the data sampling is maximized; 
ii) for the large majority of bursts the jet break has not yet occurred; 
iii) it allows an easy comparison with the X---ray luminosities 
calculated by GB05, calculated at the same time. 
For densely sampled optical
light curves, we have directly taken the flux measured at $t_{\rm
obs}=12(1+z)$ hours.  When this flux was not available, we have
interpolated between data before and after this time.
There are 2 cases (GRB 020813 and GRB 030329) in which a break in the light curve
(very likely a jet break) occurs before 12 hours.  In these cases we
have extrapolated the flux from data before the break time.

\begin{figure*}
\vskip -1 true cm
\centerline{\psfig{figure=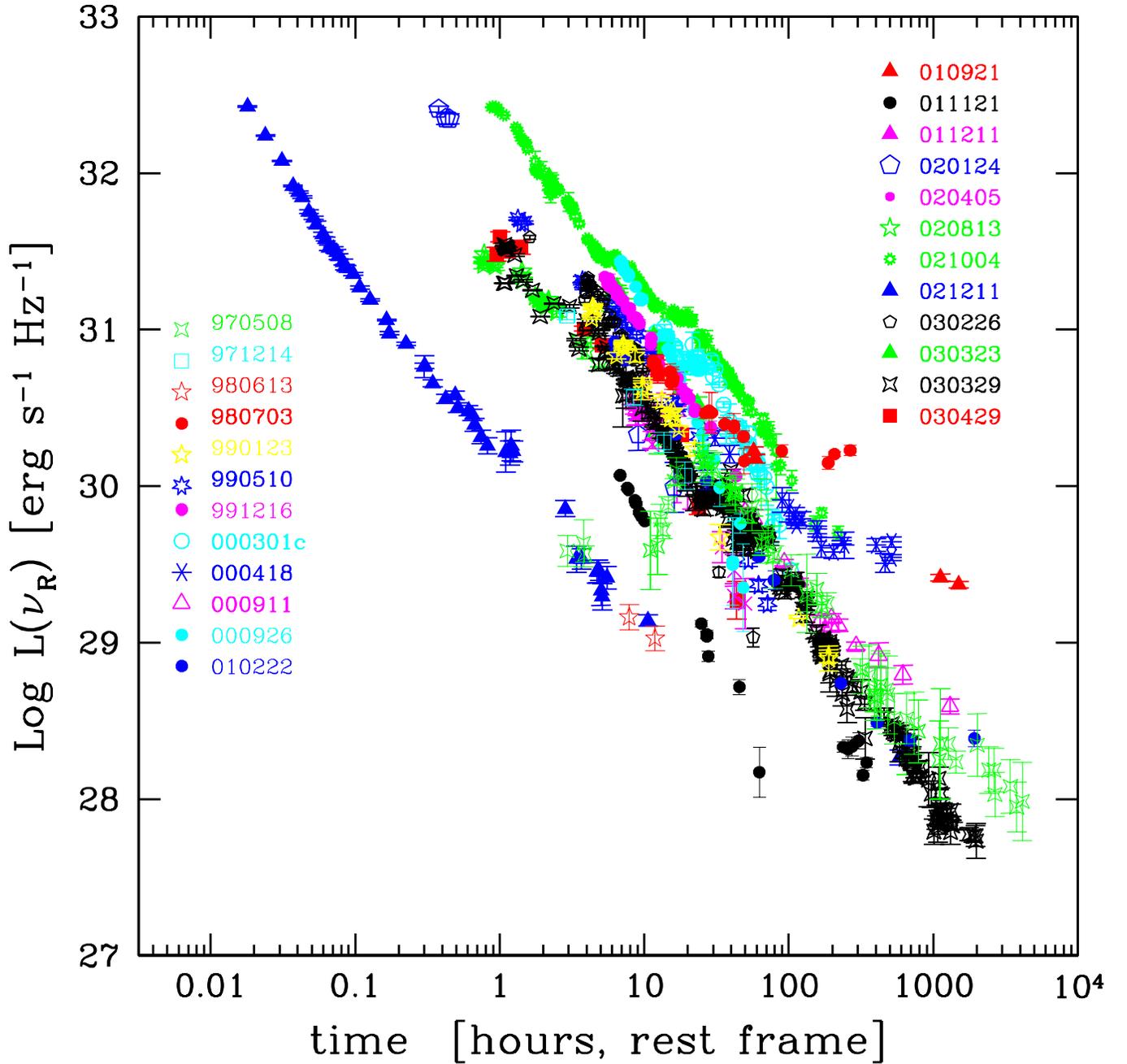,angle=0,width=21cm}}
\vskip -1.5 true cm
\caption{
Light curves of the optical luminosities as a function of the
rest frame time. The three underluminous GRBs are labeled.
All data have been corrected for extinction (both Galactic and host).
The references for the observed magnitudes can be found in the
Appendix. The references for the values of the spectral index $\beta$
and the host galaxy absorption can be found in the caption of 
Tab. \ref{tabottico}. 
} 
\label{lc1}
\end{figure*}

The observed flux $F(\nu_{\rm R}, t)$ is corrected for both galactic
and rest frame extinction.  We have calculated the $A_\lambda$ values
for the extinction in the burst host galaxies by assuming the
extinction curve of the Milky Way (Pei 1992), evaluated at the
wavelength $\lambda = 6400/(1+z)$ \AA ~(unless specifically stated
otherwise in the original reference).  
There are cases in which different authors find slightly different 
values for $A_V^{\rm host}$ and for $\beta_{o}$, i.e. the dereddened 
value of the optical spectral index.  
There is in fact some degeneracy between these two quantities
when the available data are poorly sampled and affected by relatively
large uncertainties.  
In fact, in the large majority of cases, the method used to find the 
intrinsic extinction is to assume that the spectrum is a power law, 
and the fit returns the best values of the spectral index and $A_V^{\rm host}$.
The two quantities are however somewhat correlated, since increasing 
$A_V^{\rm host}$ gives a flatter $\beta$.
In addition, different results can be obtained by using different 
extinction curves.  
Therefore, for completeness, we list in Tab. \ref{tabottico} the different 
values of $A_V^{\rm host}$ and $\beta_{o}$ found by different authors, and
the corresponding value of the optical luminosity. 
The first line of every multiple entry in Tab. \ref{tabottico}, 
corresponds to what we have used for the histograms, for Fig. \ref{lc1}
and for the following analysis.  
However, one can see that the different values of the extinction and spectral indices
do not change the derived luminosities by a large amount.
Indeed, the width of the optical luminosity distribution is not affected by
these uncertainties.
Fig. \ref{A_vhost} shows the distribution of the extinction values 
$A_V^{\rm host}$ in the host.  
Note that despite the fact that nearly half of the bursts have zero or 
almost zero host absorption, the extinction correction is crucial to obtain 
the strong clustering of the optical luminosities shown in Fig. \ref{isto_ottico}.
Without this correction, the optical luminosity distribution has a width
of $\sigma\sim 0.39$ (see Tab. \ref{sigma}).
This is partly due to those GRBs at high redshift, for which even a moderate
value of $A_V^{\rm host}$ implies a relatively large absorption at the
rest frame frequency $\nu_R (1+z)$.

%

\section{Light curves of the optical luminosities}

\begin{figure}
\vskip -0.5 true cm
\psfig{figure=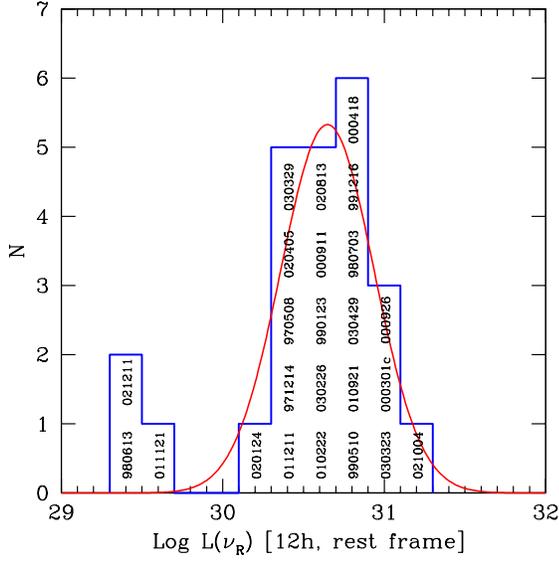,angle=0,width=8.8cm}
\vskip -0.5 true cm
\caption{ 
Histogram of the monochromatic optical luminosities 12 hours
(rest frame) after the trigger for the 24 GRBs reported in
Tab. \ref{tabottico} and shown in Fig. \ref{lc1}. 
Data have been dereddened both for galactic and host extinction. 
The solid red line represents the gaussian fit to the data
with mean value $\mu=30.65$ and dispersion $\sigma=0.28$.  }
\label{isto_ottico}
\end{figure}

In Fig.  \ref{lc1} we show the light curves in terms of the optical
luminosities in the $R$--band as a function of the rest frame time.
As can be seen, there is a clear clustering of the light curves
when corrected for the cosmological and extinction effects with respect 
to the light curves shown in Fig. 1.
Most of the luminosities at 12 h (since trigger) clusters 
around $\log L_{\nu_R}\sim$ 30.65 (see Fig. \ref{isto_ottico}).
This is the main result of our paper.
There are three exceptions, i.e.  GRB 980613, GRB 011121 and GRB 021211, 
which stand apart from the bulk of the other bursts, being underluminous by
a factor $\sim 15$ with respect to the other GRBs.

Some of the light curves shown in Fig. \ref{lc1} appear peculiar,
in particular:
\vskip 0.3 true cm

{\it GRB 970508:} the optical light
curve of this bursts showed an initial brightening followed,
approximately at 1 day, by a normal decay. 
For this reason we
calculated $L_{\nu_R}$ at 12 h by extrapolating the light curve
from the data above $\sim$30 h (rest frame).

Choosing {\it not} to extrapolate from later times would make
this burst to belong to the "underluminous family" for times
earlier than 12 hours.


{\it GRB 020813 and GRB 030329:} these GRB have an early jet break time 
(roughly at 4.6 h and 10 h, rest frame, respectively), 
and we calculate the 12 h luminosity by extrapolating from
the light curve before $t_{\rm jet}$.

Note that choosing {\it not} to extrapolate from earlier times
makes these bursts to remain in the same ``luminous burst family".

\vskip 0.3 true cm

Similarly to what we have done with the light curves
of the observed fluxes, we can derive the distribution of 
the monochromatic optical luminosities at 12 h (rest frame) for 
the bursts of Fig. \ref{lc1}. 
This is shown in Fig. \ref{isto_ottico}. 
We note the separation of the 24 GRBs into two groups: 
the bulk of GRBs (21 objects) which have a 12 h
rest frame luminosity distribution spanning less than one order of
magnitude, and a second group (3 objects) which appears underluminous
by a factor $\sim 15$.
The first distribution can be well represented by a gaussian with an average
luminosity $\langle \log L(\nu_{R},12h) \rangle=30.65$ and a dispersion
$\sigma=0.28$. 
 
The typical error on $\log L(\nu_{R},12h)$ is around 0.1, 
much less than the 1$\sigma$ dispersion of the distribution
of this quantity.
This error has been estimated by propagating the average error
on the observed magnitude (0.1), $A_V^{\rm host}$ (0.13) and $\beta$ (0.1).

We note that Bo\"er \& Gendre (2000) have analyzed the behavior of the
optical afterglow of the bursts studied in their paper (8 in total),
without applying the dereddening of
the extinction of the host (at that time largely unavailable). 
They did not find any clustering, nor a dichotomy, although
(even without correcting for the absorption of the host), 
they noted that the distribution of the intrinsic optical 
luminosities was narrower than the distribution of the observed fluxes.

Note that the choice of 12 hours rest frame is not critical
for our results, as can be seen in Fig. \ref{lc1}, as long as
the chosen time is less than the jet break time
for most bursts.

Our result is surprising in many respects, as mentioned in the
introduction.
We can compare this narrow clustering of the optical luminosity with
the distribution of the prompt emission isotropic energy 
$E_{\rm \gamma, iso}$ for the same bursts (see Tab. 4). 
The $\log E_{\rm \gamma, iso}$ distribution (if fitted with a gaussian) 
has a much larger dispersion of $\sigma\sim 0.8$ (see Tab. \ref{sigma}).
Another unexpected result concerns the distribution of the ratio
of $\log[L(\nu_{R},12h)/E_{\rm \gamma, iso}]$.
Since we expect that the afterglow luminosity depends
upon the isotropic kinetic energy of the fireball,
which should be measured by $E_{\rm \gamma, iso}$,
we naively expect this distribution to have a smaller dispersion 
than either the $\log L(\nu_{R})$ or the $\log E_{\rm \gamma,iso}$ 
distribution.
Instead, the $\log[L(\nu_{R},12h)/E_{\rm \gamma, iso}]$
distribution has a dispersion $\sigma=0.9$, even larger than the
dispersion of $\log E_{\rm \gamma, iso}$.

The three underluminous bursts, (i.e.  GRB~980613, 011121, 021211)
which seem to form a separate ``family" are more than 4 $\sigma$
dimmer than the majority of bursts.
Note that for GRB~011121 and GRB~021211 the two
parameters $\beta_o$ and $A_{V}^{\rm host}$ have possible different
estimates (see Tab. \ref{tabottico}). 
Here we adopted the values reported in the 
first line of Tab. \ref{tabottico}. 
However, if we consider the other possible choices, the implied
$L(\nu_{R},12h)$ would be even smaller, making these two bursts more
inconsistent (more than 4.5 $\sigma$) with the distribution of the bulk
of the other GRBs.  
Instead, for what concerns the GRBs with
different estimates of $\beta_o$ and $A_{V}^{\rm host}$ that fall in
the more populated group, we note that using the other choices
would shift their luminosities by less than one $\sigma$
(except for GRB 980703, for which the shift would be 1.7 $\sigma$).

\begin{figure}
\vskip -0.5 true cm
\psfig{figure=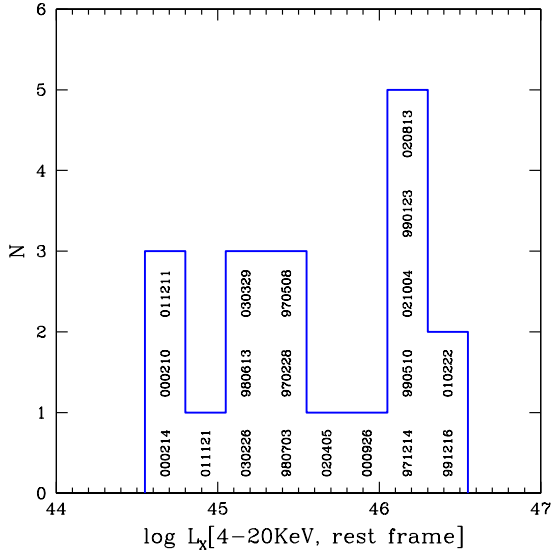,angle=0,width=8.8cm}
\vskip -0.5 true cm
\caption{
Histogram of the X--ray luminosities 12 hours 
(rest frame) after the trigger, calculated in the 
rest frame band [4--20 keV].
}
\label{isto_x}
\end{figure}

The three underluminous bursts do not seem to have any
distinguishing property other than their smaller optical
luminosities: all three have ``normal" optical decays, 
spectral indices and extinction values.
However, note that 2 of these GRBs lies on the faint portion of 
the X--ray luminosity distribution, see Fig. \ref{isto_x} (for the third there
is no X--ray detected afterglow).
On the other hand, the faint end of the X--ray luminosity distribution 
contains also bursts which belong to the ``bright" optical family
(i.e. GRB 011211 and GRB 030326).
Note also that GRB 000210 and GRB 000214, lying at the faint extreme
of the X--ray luminosity distribution, are dark bursts.

\begin{table}
\begin{center}
\begin{tabular}{l|l}
\hline\hline 
Distribution & $\sigma$ \\ 
\hline 
$\log L(\nu_{R})$ @ 12h rest frame & 0.28$^a$ \\ 
$\log L_X$ [4--20 keV], @12h rest frame & 0.74$^b$ \\ 
$\log F(\nu_{R})$ @ 12h obs frame & 0.48 \\
$\log L(\nu_{R})$ @ 12h rest frame, no $A_V^{\rm host}$ & 0.39$^{a,c}$\\
$\log E_{\rm \gamma, iso}$ & 0.80 \\ 
$\log [\nu_{R}F(\nu_{R})t_{12h}$/Fluence$_{\gamma, {\rm
      iso}}]$ & 0.93 \\  
$\log [\nu_{R}L(\nu_{R})t_{12h}/E_{\rm \gamma, iso}]$ & 0.9 \\ 
\hline
\end{tabular}
\caption{Width of the distributions of different quantities, 
according to a Gaussian fit.  
$a$: considering all bursts but the 3 underluminous ones.  
$b$: formal result from the fit, but the fit is poor. 
$c$: optical luminosities have been dereddened only for galactic
absorption, no host galaxy extinction has been considered.}
\label{sigma}
\end{center}
\end{table}

\begin{figure}
\vskip -0.5 true cm
\centerline{\psfig{figure=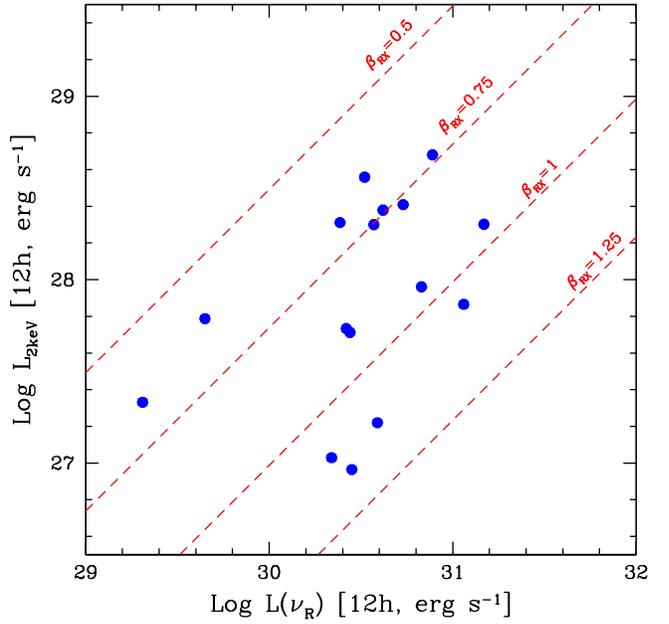,angle=0,width=10cm}}
\vskip -0.9 true cm
\caption{
X--ray monochromatic [2 keV, rest frame] luminosity 
as a function of the optical monochromatic 
[$R$ band, rest frame] luminosity at 12 hours after trigger.
The dashed lines correspond to different broad band spectral
indices $\beta_{RX}$ as labelled.
}
\label{lxlo}
\end{figure}
\begin{figure}
\vskip -0.5 true cm
\centerline{\psfig{figure=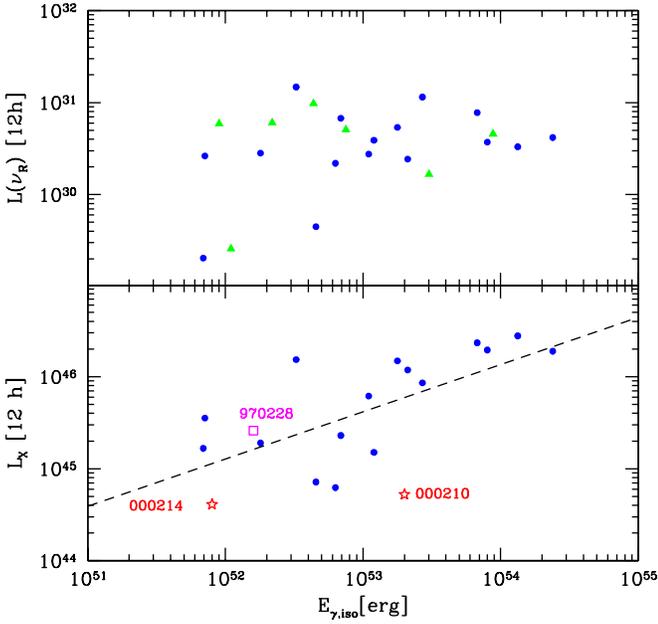,angle=0,width=10cm}}
\vskip -0.9 true cm
\caption{
Optical monochromatic luminosity (top panel)
and X--ray [4--20 keV] luminosities (mid panel)
at 12 hours after trigger (rest frame time)
as a function of the isotropic emitted energy during the 
prompt phase (integrated between 1 keV and 10 MeV, see Ghirlanda,
Ghisellini \& Lazzati 2004).
Circles corresponds to bursts having both optical and X--ray data.
Triangles are GRBs with optical but no X--ray data.
Stars are the two GRBs (as labelled) with X--rays but no
optical data.
We also show GRB 970228, for which there is no information on
the amount of extinction in the host (square).
The dashed line is the linear regression fit
($\log L_x \propto 0.51 \log E_{\gamma, \rm iso}$),
which has a chance probability $P=3\times 10^{-3}$.
}
\label{l_eiso}
\end{figure}
\begin{figure}
\vskip -0.5 true cm
\centerline{\psfig{figure=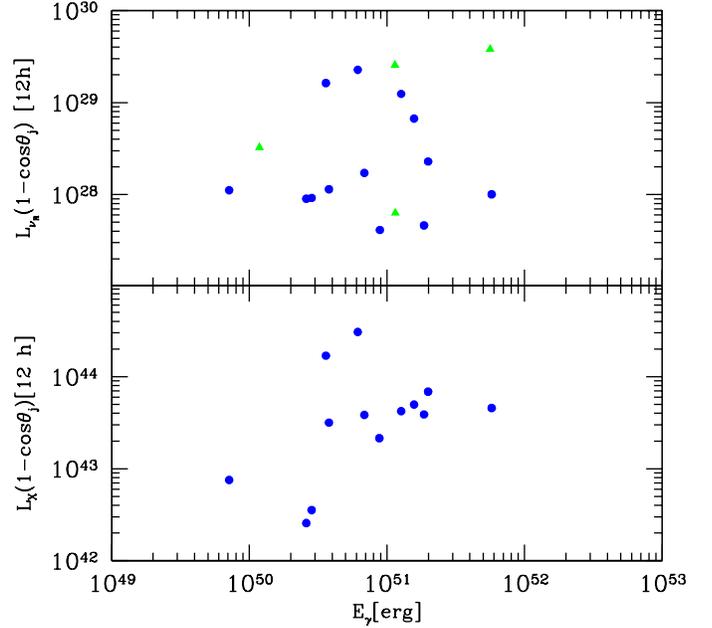,angle=0,width=10cm}}
\vskip -0.9 true cm
\caption{
For those bursts of measured $\theta_{\rm j}$, 
we have calculated the collimation corrected 
afterglow luminosity (top panel: optical; bottom
panel: X--ray) 
at 12 hours after trigger (rest frame time)
as a function of the collimation corrected emitted energy during the 
prompt phase (integrated between 1 keV and 10 MeV, see Ghirlanda,
Ghisellini \& Lazzati 2004).
Symbols as in Fig. \ref{l_eiso}
Note that there is now no correlation between the optical or the 
X--ray luminosity and the prompt emitted energy.
}
\label{l_egamma}
\end{figure}




\begin{table*}
\begin{center}
\begin{tabular}{l l c c c c c c c}
\hline\hline
GRB &$z$ &$\alpha$ & $\beta_X$ & $F_X$         & $t_{\rm obs}$ & band &
$\log L_{10~{\rm keV}}^{12h}$ &Ref\\ 
    &    &         &           &$10^{-12}$ cgs &             & keV  & & \\
\hline
\object{970228} &0.695$^1$ &1.3\pim0.1   &1.1\pim0.3   &1.0\pim0.2    &1d   &2--10
  &26.81 & co97 \\  
970508 &0.835 &1.1\pim0.1$^a$ &1.1\pim0.3   &1.0\pim0.4    &1d
  &2--10 &26.97 & gb05,pi98\\ 
971214 &3.418 &1.1$\pm$0.1  &1.2$\pm$0.4  &0.23$\pm$0.05 &1d   &2--10
  &27.47 & co99, gb05\\ 
980613 &1.096 &1.1\pim0.2   &1            &0.27\pim0.07  &1d   &2--10
  &26.63 & co99\\   
980703 &0.966 &0.9\pim0.2   &1.8\pim0.4   &0.48\pim0.07  &1d   &2--10
  &26.70 & gb05\\ 
990123 &1.60  &1.35         &0.99\pim0.07 &5.3\pim0.2    &11h &1.6--10
  &27.69& dp03,he99 \\  
       &      &1.44\pim0.11 &1.00\pim0.05 &1.8\pim0.4    &1d   &2--10
       & 27.72 &gb05 \\  
990510 &1.619 &1.4\pim0.1   &1.2\pim0.2   &1.2\pim0.2    &1d   &2--10
  &27.57 & gb05\\  
991216 &1.02  &1.6\pim0.1   &1.2\pim0.2   &2.58          &37h &2--10
  &27.77 & pi00 \\
  &      &      1.6\pim0.1 &0.8\pim0.5   &5.6\pim0.3    &1d   &2--10
  &27.89 & gb05 \\ 
\object{000210} &0.846$^2$ &1.38\pim0.03 &0.95\pim0.15 &0.4\pim0.06   &11h &2--10
       &26.13 & pi02 \\ 
       &      &1.38\pim0.03 &0.9\pim0.2   &0.21\pim0.06  &1d   &2--10
       &26.32 & gb05 \\ 
\object{000214} & 0.42$^3$ &0.8\pim0.5$^b$ &1.2\pim0.3 &0.77\pim0.08  &15h &2--10
        &26.01 & an00\\ 
       &      &0.7\pim0.3   &1.2\pim0.5   &0.6\pim0.2    &1d   &2--10
    &26.05 & gb05\\  
000926 &2.066 &1.7\pim0.5   &0.7\pim0.2   &0.12\pim0.1   &2.78d &2--10
   &27.35 & gb05\\   
010222 &1.477 &1.33\pim0.04 &1.01\pim0.06 &2.7\pim0.6    &1d   &2--10
   &27.85 & gb05 \\  
011121 &0.36  &$4^{+3}_{-2}$&2.4\pim0.4   &0.6\pim0.2    &1d   &2--10
  &26.11 & gb05 \\ 
011211 &2.14  &1.3\pim0.1  &1.2\pim0.1    &0.03\pim0.01  &1d   &2--10
  &26.19 & gb05 \\ 
020405 &0.69  &1.9\pim1.1  &0.72\pim0.21  &1.36\pim0.25  &1.71d
  &0.2--10 &27.21 & mi03\\   
020813 &1.25  &1.38\pim0.06&0.85\pim0.04  &2.2           &1.33d
  &0.6--6 &27.70 & bu03 \\  
021004 &2.33  &0.9\pim0.1  &1.01\pim0.08  &0.63          & 1.37d
  &0.6--6 &27.60 & but03 \\  
030226 &1.98  &2.7\pim1.6  &0.9\pim0.2    &0.035\pim0.002&1.77d &2--10
  &26.59 & gb05\\ 
030329 &0.168 &0.9\pim0.3  &0.9\pim0.2    &14.3\pim2.9   &1d   &2--10
  &26.69 & gb05  \\ 
\hline 
\end{tabular}
\caption{
X--ray properties of the GRBs with known redshift.
Data have been collected from the literature. 
$\alpha$ and $\beta_X$ are the temporal and spectral
power law indices, respectively 
[i.e. $F(\nu,t)\propto t^{-\alpha}\nu^{-\beta_X} $]. 
$F_X$ is the observed X--ray flux integrated in the
reported energy band and $L_{10 \rm keV}^{12h}$ is the monochromatic 
X--ray luminosity at 12 h (rest frame) calculated at 10 keV.  
$a$: 970508 showed a substantial rebrightening, correlated
with the optical (Piro et al. 1998). $b$: decay index considering only
MECS data. If WFC is included, $\alpha=1.41\pm 0.03$ (Antonelli et
al. 2000).  For GRB 030329 we have calculated the X--ray flux at 12 hours
rest frame extrapolating from earlier data, since this GRBs showed a jet
break at approximately 10 hours (rest frame time) (see Tiengo et al. 2004).
References: 
co97: Costa et al., 1997; 
gb05: Gendre \& Bo\"er, 2005; 
pi98: Piro et al., 1998; st04: 
Stratta et al., 2004; 
co99: Costa et al., 1999; 
dp03: De Pasquale et al. 2003; 
he99: Heise et al., 1999; 
pi00: Piro et al., 2000; 
pi02: Piro et al., 2002; 
an00: Antonelli et al. 2000; 
mi03: Mirabal et al., 2003; 
bu03: Butler et al., 2003.
1) Djorgovski et al. 1999;   
2) Piro et al. 2002;
3) Antonelli et al. 2000.
}
\end{center}
\label{tabx}
\end{table*}

\subsection{Comparison with X--ray afterglow luminosities}

GB05, studying the X--ray afterglow light curves, found that the
distribution of the rest frame [4--20] keV
\footnote{The light curves plotted in their Fig. 2 refer to the [2--10]
keV band, but they locate all bursts at $z=1$.  Therefore the rest
frame energy band is 4--20 keV.}  
X--ray luminosities is bimodal, clustering around two values.  
We have expanded the original list of GB05 by the inclusion 
of three more GRBs: GRB 020405, GRB 020813, GRB 021004.  
We have also modified slightly some of the data presented in
their original table, as new information is now available for some of
the bursts.  
For this reason we have collected in Tab. \ref{tabx} the
information about the X--ray data, with the appropriate references.
With respect to the results obtained by GB05, we find a more continuous 
distribution, as can be seen in Fig. \ref{isto_x}, without a 
clear clustering or a clear separation in two GRB ``families".

Fig. \ref{isto_x} shows that the distribution of X--ray luminosities
is wider than the distribution of the optical luminosities.  
A gaussian fit (although poor) 
gives a dispersion $\sigma=0.74$ (see Tab. \ref{sigma}).

Fig. \ref{lxlo} shows the monochromatic [2 keV, rest frame] 
X--ray luminosity as a function of the optical monochromatic 
luminosity [$R$--band] 12 hours after trigger, in the rest frame.  
Dashed diagonal lines correspond to lines of constant broad band
spectral indices $\beta_{RX}$ between the optical ($R$--band) and the
X--ray (2 keV), defined as
\begin{equation}
\beta_{RX}\, =\, { \log (L_{\nu_R}/L_X) \over \log(\nu_X/\nu_R)}
\end{equation}
This figure shows that bursts that
are more luminous in X--rays tend to have flatter $\beta_{RX}$
spectral indices (and therefore they are relatively less luminous in
the optical) and viceversa.  
There are two exceptions, both belonging
to the dim optical family.  
This behavior (flatter $\beta_{RX}$ for greater $L_X$)
is a necessary condition for having optical luminosities
more clustered than the X--ray ones.

\subsection{Comparison with the emitted $\gamma$--ray energies}

\begin{table}
\begin{center}
\begin{tabular}{l l l l}
\hline\hline
GRB    & $z$ & $E_{\rm \gamma, iso}$ & $\theta_{\rm j}$ \\
       &     & [erg]                 & [deg]    \\
\hline
970508 & 0.835 & 7.1E51  (0.15) & 24.0 (3.3)\\
971214 & 3.418 & 2.11E53 (0.24) & ...   \\
980613 & 1.096 & 6.9E51  (0.95) & ...    \\
980703 & 0.966 & 6.9E52  (0.82) & 11   (0.8) \\
990123 & 1.6   & 2.39E54 (0.28) & 3.98 (0.57)\\
990510 &1.616  & 1.78E53 (0.19) & 3.74 (0.24)\\
991216 & 1.02  & 6.7E53  (0.81) & 4.4  (0.6)\\
000301c& 2.067 & 4.37E52        & 13.14 \\
000418 & 1.1181& 7.51E52        & 22.3 \\
000911 & 1.06  & 8.8E53  (1.05) & ... \\ 
000926 & 2.0375& 2.7E53         & 6.19 \\
010222 & 1.477 & 1.33E54 (0.15) & 3.03 (0.14)\\
010921 & 0.45  & 9.0E51  (1.0)  & ...      \\
011121 & 0.36  & 4.55E52 (0.54) & ...      \\
011211 & 2.14  & 6.3E52  (0.7)  & 5.2  (0.63)\\
020124 & 3.198 & 3.02E53 (0.36) & 5.0  (0.3)\\
020405 & 0.69  & 1.1E53  (0.13) & 6.4  (1.05)\\
020813 & 1.25  & 8.0E53  (0.96) & 2.7  (0.13)\\
021004 & 2.3351& 3.27e52 (0.4)  & 8.5  (1.04)\\
021211 & 1.004 & 1.1E52  (0.13) & ...          \\
030226 & 1.986 & 1.2E53  (0.13) & 3.94 (0.49)\\
030329 & 0.1685& 1.8E52  (0.21) & 5.1  (0.4) \\
030429 & 2.66  & 2.19E52 (0.26) & 5.96 (1.43)\\
\hline
\end{tabular}
\caption{Values of redshift, $E_{\rm \gamma, iso}$ and the
semiaperture jet angle $\theta_{\rm j}$ used in Fig. \ref{l_eiso}.
Values taken from Ghirlanda et al. 2004.
When present, the values in parenthesis are the errors.}
\end{center}
\label{tab_eiso}
\end{table}

Fig. \ref{l_eiso} shows both the optical and the X--ray luminosities
as a function of $E_{\rm \gamma, iso}$, the isotropic energy emitted
during the prompt phase.  There is no correlation in the case of the
optical luminosities (top panel), while there is some correlation in
the case of X--ray luminosities, albeit not very strong (chance
probability $P\sim 3\times 10^{-3}$).  Both quantities ($L_x$ and
$E_{\rm \gamma, iso}$), being isotropic quantities, depend on the aperture
angle of the jet $\theta_{\rm j}$.  For those GRBs for which we know
$\theta_{\rm j}$ (13 objects) we can construct the collimation
corrected quantities by multiplying by $(1-\cos\theta_{\rm j})$.
After doing that, the correlation between the X--ray luminosity and
the $\gamma$--ray energy disappears (bottom panel of
Fig. \ref{l_egamma}).

We would have expected the same effect in the case of
the optical luminosities, i.e. there should be an apparent correlation
when considering isotropic quantities, but there is not (upper panel
of Fig. \ref{l_eiso}).
Furthermore, the distribution of the collimation corrected optical
luminosities becomes not narrower, but slightly {\it broader} than
the distribution of the isotropic luminosities,
as can be seen from the upper panel of Fig. \ref{l_egamma}, which
also shows that there is no correlation between the collimation
corrected optical luminosities and prompt emitted energies.
We have also verified that 
there is no correlation between the optical or the X--ray luminosity
and the spectral peak energy $E_{\rm peak}$ of the prompt emission.




%
\begin{figure*}
\vskip -0.5 true cm
\psfig{figure=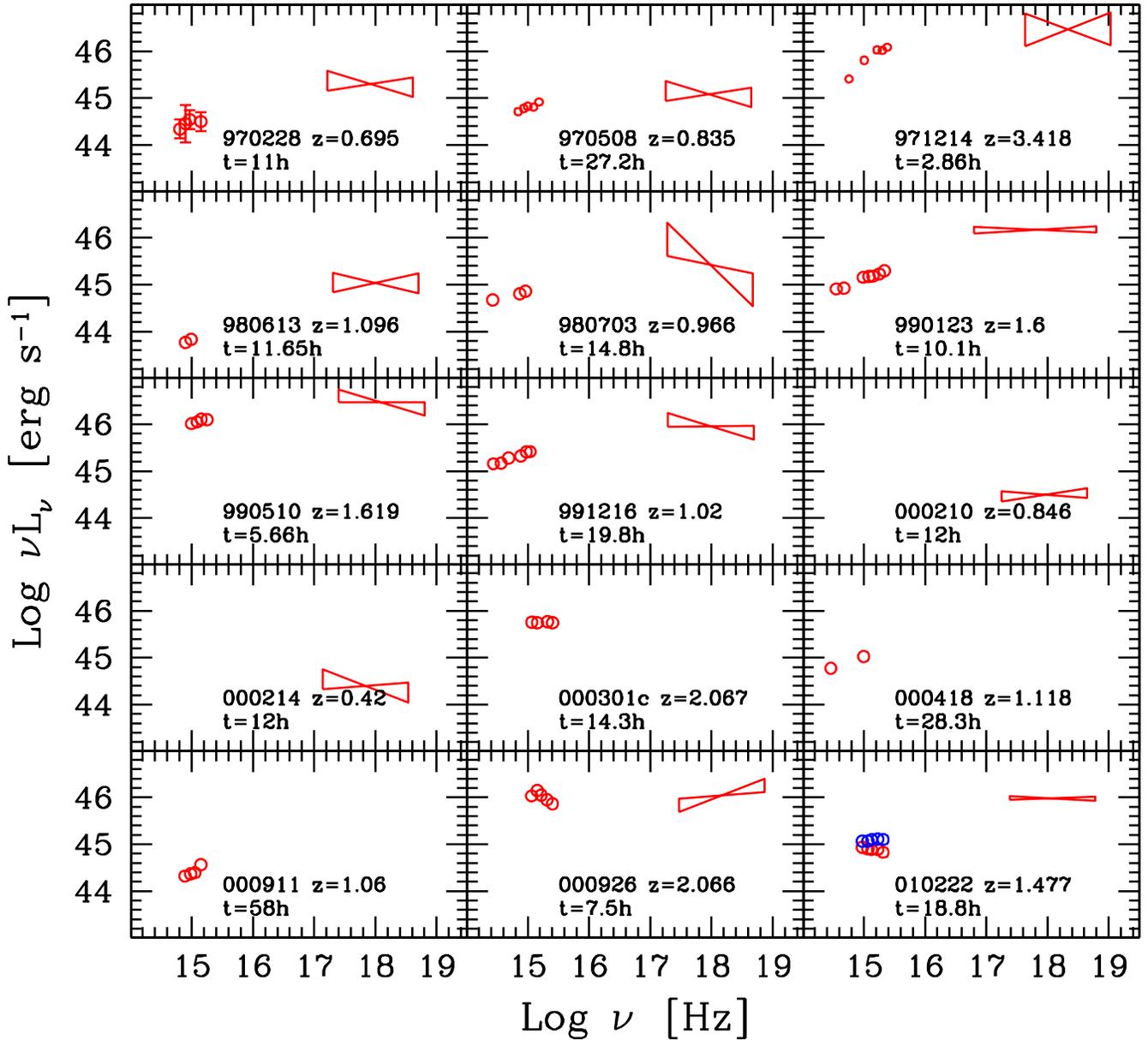,angle=0,width=19cm}
\vskip -1.5 true cm
\caption{
Optical to X--ray spectral energy distribution for all GRBs 
in our sample. 
Data are simultaneous, at the rest frame time labelled in each panel. 
Sources of data:
GRB 970228: optical: Reichart 1999; X--rays: Costa et al. 1997.
GRB 970508: Galama et al. 1998; Piro et al. 1998.
GRB 971214: Wijers et al. 1999; Stratta et al. 2004.
GRB 980613: Hjorth et al. 2002; Costa 1999.
GRB 980703: Vreeswijk et al. 1999; De Pasquale et al. 2003.
GRB 990123: Galama et al. 1999; Heise et al. 1999.
GRB 990510: optical: Harrison et al. 1999; X--rays: Gendre \& Bo\"er 2005. 
GRB 991216: Halpern et al. 2000, Garnavich et al. 2000; Piro et al. 2000.
GRB 000210: Piro et al. 2002.
GRB 000214: Antonelli et al. 2000.
GRB 000301c: Jensen et al. 2001.
GRB 000418: Klose et al. 2000.
GRB 000911: Masetti et al. 2005.
GRB 000926: optical: Fynbo et al. 2001.; X--rays: Gendre \& Bo\"er 2005.
GRB 010222: optical: Masetti et al. 2001.; X--rays: Gendre \& Bo\"er 2005.
}
\label{sed1}
\end{figure*}
\begin{figure*}
\vskip -0.5 true cm
\psfig{figure=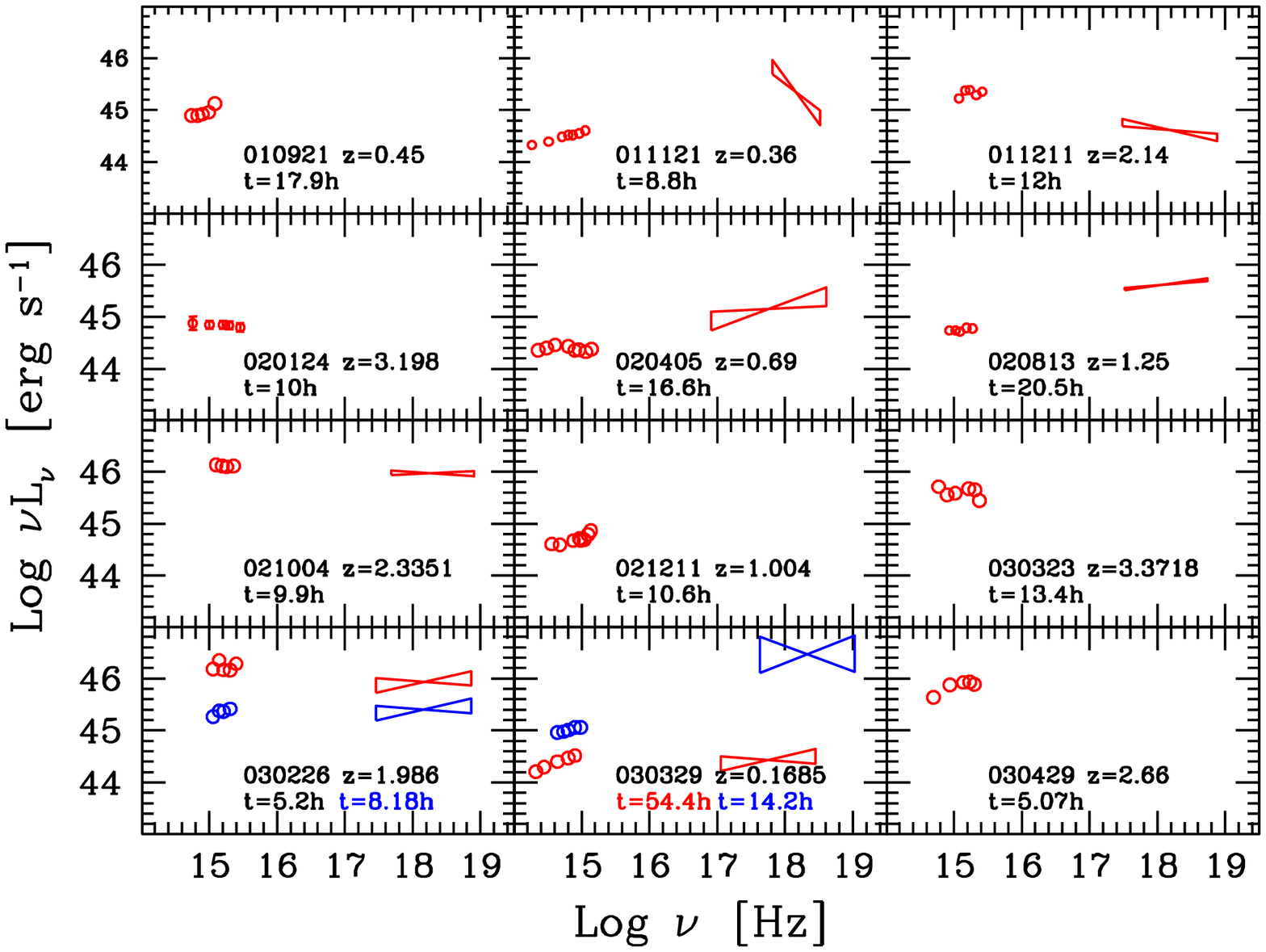,angle=0,width=19cm}
\vskip -4.5  true cm
\caption{
Optical to X--ray spectral energy distribution for all GRBs 
in our sample. 
Data are simultaneous, at the rest frame time labelled in each panel. 
Sources of data:
GRB 010921: Price et al. 2002.
GRB 011121: optical: Garnavich et al. 2003; X--rays: Gendre \& Bo\"er
2005. 
GRB 020124: Hjorth et al. 2003.
GRB 020405: Masetti et al. 2003; Mirabal et al. 2003.
GRB 020813: optical: Covino et al. 2003;  X--rays: Butler et al. 2003.
GRB 021004: optical: Pandey et al. 2003;  X--rays: Butler et al. 2003.
GRB 021211: Nysewander et al. 2005.  
GRB 030323: Vreeswijk et al. 2004.
GRB 030226: optical: Pandey et al. 2004; X--rays: Gendre \& Bo\"er
2005.   
GRB 030329: optical: Bloom et al. 2004; Matheson et al. 2003; X--rays:
Gendre \& Bo\"er 2005.
GRB 030429: Jakobsson et al. 2004. 
}
\label{sed2}
\end{figure*}

\section{Spectral Energy Distributions}

Fig. \ref{sed1} and Fig. \ref{sed2} show the optical to 
X--ray Spectral Energy Distribution (SED) for all bursts in our sample.
Data are plotted in the rest frame of the source,
after being corrected for extinction.
In constructing the SED we have considered the optical multiband
photometry at a time as consistent as possible with the
X--ray observations, requiring the least possible
extrapolation from data taken at other times. 
In some cases (i.e. GRB 030226 and GRB 030329) we plotted
two SED for each bursts corresponding to two different observing
times or, in the case of GRB 010222, corresponding to two different
choices of host galaxy optical extinction. 
The captions of these figures report the original source of data.

This figure shows that in a large fraction of cases 
the optical to X--ray data seem to be consistent with 
being produced by the same emission process.
In a minority of cases (GRB 000926, possibly GRB 020405 and 
the earlier SED of GRB 030226) the optical spectrum
is steeper than the X--ray spectrum, suggesting that they are
produced by a different component.
One obvious possibility is the inverse Compton process dominating
the X--ray flux at the observing times.
There is finally one case (GRB 020813) where the optical and the
X--ray emission 
smoothly join, but the peak of the overall SED lies above the X--ray band.
To summarize:

\begin{itemize}

\item Of the 27 SED of GRBs shown, 17 have both optical and X--rays
 (note that we now include GRB 970228),
 2 have only the X--ray data (GRB 000210 and GRB 000214) and 8 have only
 the optical data (GRB 000301c; GRB 000418; GRB 000911; GRB 010921; GRB 020124;
 GRB 021211; GRB 030323; GRB 030429).
 
\item In 13 out of 17 cases, the extrapolated optical and X--ray spectra join 
  smoothly, indicating a common (synchrotron) origin by a population
  of electrons 
  characterized by an energy break (flatter at low energies and steeper at high
  energies, as expected in the case of incomplete cooling).
 
\item Of the remaining 4, GRB 000926 shows a steep optical and a flat
 X--ray spectrum, 
 suggesting that the X--ray flux has a non--synchrotron origin.
 The same (but less extreme) behavior characterizes the SED of GRB
 020405 and the early 
 time SED of GRB 030226. The SED of GRB 010222 is somewhat difficult
 to classify, since 
 the optical spectrum could smoothly join the X--ray one if the absorption is
 slightly underestimated.
 Note that GRB 000926 and GRB 010222 are the two bursts lying in between the 
 two groups of X--ray luminosities identified by GB05.

\item Of the 13 ``normal synchrotron" SEDs, in 11 cases the $\nu L_\nu$ peak,
 $\nu_{\rm peak}$, is between the optical and the X--ray band or in
 the X--ray band.  
 The uncertain cases are due to the relatively large uncertainties of
 the X--ray slope, 
 which are often characterized by a spectral index close to unity 
(i.e. flat in $\nu L_\nu$).
 In GRB 021004 $\nu_{\rm peak}$ could be in the IR band, but the overall spectrum
 is nearly flat in $\nu L_\nu$.
 In GRB 020813 $\nu_{\rm peak}$ is above the X--ray range.

\end{itemize}

\begin{table}[h!]
\begin{center}
\begin{tabular}{lllll}
\hline
\hline
GRB  &$z$ &$A_R^{\rm Gal}$ & $\log L(\nu_R)$ &$\log L_X$ \\ 
\hline
\object{050126} &1.29   & 0.15  &            & 44.8   \\ 
\object{050315} &1.949  & 0.13  & 30.33      & 47.0    \\ 
\object{050318} &1.44   & 0.046 &            & 45.7  \\ 
\object{050319} &3.24   & 0.031 & 30.7       & 46.4  \\ 
\object{050401} &2.90   & 0.175 & 30.4$^a$   & 46.9 \\ 
\object{050416} &0.654  & 0.13   &           & 44.9  \\ 
\object{050505} &4.3    & 0.058  & 30.5      & 46.3            \\ 
\object{050525} &0.606  & 0.255  & 29.6$^a$  & 45.4       \\ 
\object{050603} &2.821  & 0.074  & 30.8  & 46.0         \\ 
\object{050730} & 3.967 & 0.135  & 30.79$^a$     & 45.9     \\ 
\object{050820A} & 2.612 & 0.123 & 30.81    & 46.65  \\ 
\object{050824}  & 0.83 & 0.093 & 29.64 &\\
\object{050904}  & 6.29 & 0.161 & 31.18$^b$ &\\
\object{050908} &3.3437 & 0.069& 30.52 &\\
\object{050922C}& 2.198 & 0.27 & 30.34 &\\
\object{051016B}& 0.9364 & 0.13 & & \\
\object{051109} & 2.346 & 0.508 & 30.63 &\\
\object{051111} & 1.55 & 0.43 & 30.24& \\ 
\hline
\end{tabular}
\caption{
SWIFT long bursts with spectroscopically measured redshift.
For those 
with enough photometric optical data we estimate
their luminosity at 12 hours from trigger (rest frame). 
We assumed $\beta_{opt}=1$. All these luminosities but three
have been dereddened for the galactic extinction, but not for the
(still unknown) extinction in the host galaxy.
$a$: corrected for the host galaxy absorpion and using the spectral
index $\beta_o$ found in litterature; 
$b$: extrapolated from J band photometry.
References:
GRB 050315: Cobb et al., 2005, GCN 3104; Cobb et al., 2005,
GCN 3110, z: Kelson et al. 2005, GCN 3101.
GRB 050319: Wo\'zniak et al., 2005, z: Fynbo et al., 2005b GCN 3136; 
GRB 050401: McNaught et al., 2005, GCN 3163; D'Avanzo et al.,
2005, GCN 3171; Kahharov et al., 2005, GCN 3174; Misra et al.,
2005, GCN 3175; Greco et al., 2005, GCN 3319; Watson et al., 2005, z:
Fynbo et al., 2005c GCN 3176.
GRB 050505: Rol et al., 2005, GCN 3372; Chapman et al., 2005, 
GCN 3375; Klotz  et al., 2005, GCN 3403, z:
Berger et al., 2005, GCN 3368.
GRB 050525: Blustin et al. 2005, z: Foley et al. 2005b GCN 3483.
GRB 050603: Brown et al., 2005, GCN 3549, z: berger et al., 2005c GCN 3520.
GRB 050730: Sota et al., 2005, GCN 3705; Holman et al., 2005, GCN 3716,
Haislip et al., 2005, GCN 3719; Klotz et al., 2005, GCN 3720, z:
Holman et al., 2005, GCN 3716. 
GRB 050820: Prochaska et al., 2005; Fox \& Cenko 2005, GCN 3829;
Cenko \& Fox, 2005, GCN 3834; Page et al., 2005, GCN 3837, z: Ledoux
et al., 2005, GCN 3860.
GRB 050824: Gorosabel, J, 2005 GCN 3865; Halpern, J., P., GCN 3907, z:
Fynbo et al., 2005, GCN 3874.
GRB 050904: Haislip, J.B., et al. 2005, GCN 3914; Tagliaferri, G., et
al., 2005, z: Kawai et al., 2005, GCN 3937.
GRB 050908: Torii, K., 2005, GCN 3943; Kirschbrown, J., 2005, GCN
  3947, Foley, R.J.,, 2005, GCN 3949; Durig, D., T., 2005, GCN 3950 z:
  Fugazza et al., 2005, GCN 3948. 
GRB 050922C: Fynbo, J.P.U., 2005, GCN 4040; Durig, D.T., 2005, GCN
  4023; Covino, S., 2005, GCN 4046, z: D'Elia et al., 2005, GCN 4044; 
GRB 051109: Mirabal, N., et al., 2005,GCN 4215; Milne, P.A., et al.,
  2005, GCN 4218; Jelinek, M., et al., 2005, GCN 4227; Huang, F.Y., et
  al., 2005, z: Quimby et al., 2005, GCN 4221.
GRB 051111:  Mundell, C.G., et al., 2005, GCN 4250; Milne, P.A., et al.,
  2005, GCN 4252; Garimella, K., et al., 2005, GCN 4257; Huang, F.Y., et
  al., 2005, GCN 4258; Cameron, P.B., et al., 2005, GCN 4266, z: Hill
  et al., 2005, GCN 4255.
}  
\label{swift}
\end{center}
\end{table}

\begin{figure}
\vskip -0.5 true cm
\psfig{figure=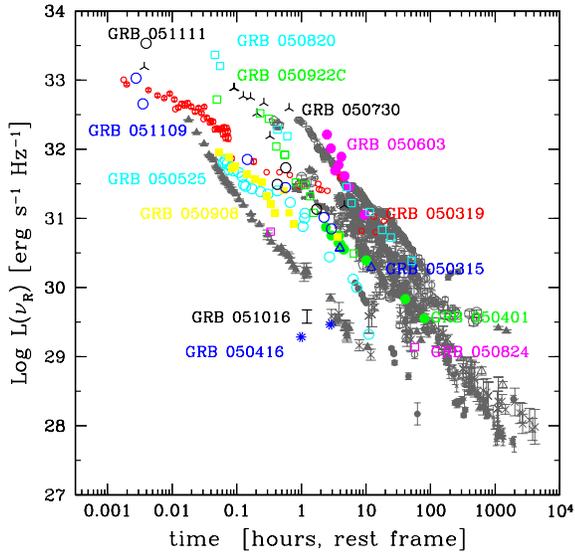,angle=0,width=8.8cm}
\vskip -0.5 true cm
\caption{In this figure we superposed the light curves of the optical
  luminosities in the rest frame of the 14 GRBs detected by SWIFT with
  enough photometric data to the light curves already shown in figure
  \ref{lc1} (grey dots). 
  The luminosities of all SWIFT GRBs are corrected only for the     
  galactic absorption,  except for GRB 050401, for which $A_V^{\rm
  host}=0.67$ (Watson et al., 2005), GRB 050525, for which 
  $A_V^{\rm host}=0.25$\pim$0.15$ (Blustin et al. 2005) and
  GRB 050730 for which $A_V^{\rm host}\sim 0$ (Starling et al. 2005).
  The SWIFT GRBs luminosities have been k--corrected assuming a common 
  spectral index $\beta=1$. The names of
  the SWIFT GRBs are given near their light curves.
}
\label{luce_swift}
\end{figure}

\begin{figure}
\vskip -0.5 true cm
\psfig{figure=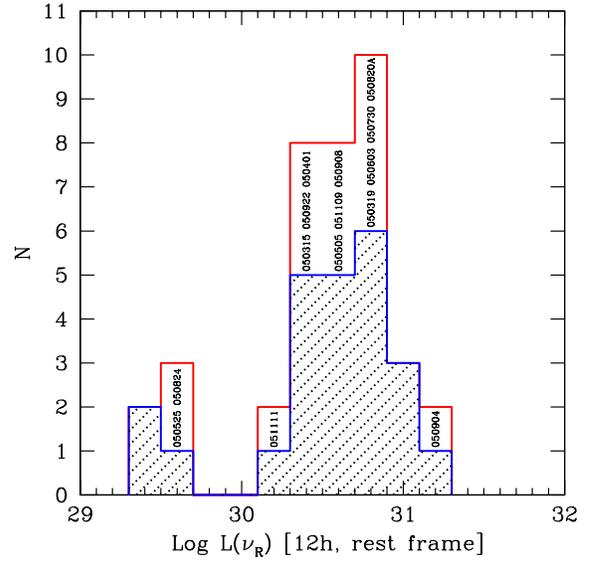,angle=0,width=8.8cm}
\vskip -0.5 true cm
\caption{
In this histogram we added (to the histogram shown in Fig.
  \ref{isto_ottico}), the monochromatic optical
  luminosities 12 hours (rest frame) after the trigger of the 14 GRBs
  discovered by SWIFT whose intrinsic luminosities could be calculated
  from the  photometric data.
  We caution the reader that the optical luminosities
  of all SWIFT bursts (except GRB 050401, GRB 050525 and GRB 050730)
  are uncorrected   
  for the host galaxy absorption and are $k$--corrected assuming a 
  the same optical spectral index ($\beta=1$) for all bursts
  (except GRB 050401, GRB 0505025 and GRB 050730).
} 
\label{isto_swift}
\end{figure}

\begin{figure}
\vskip -0.5 true cm
\psfig{figure=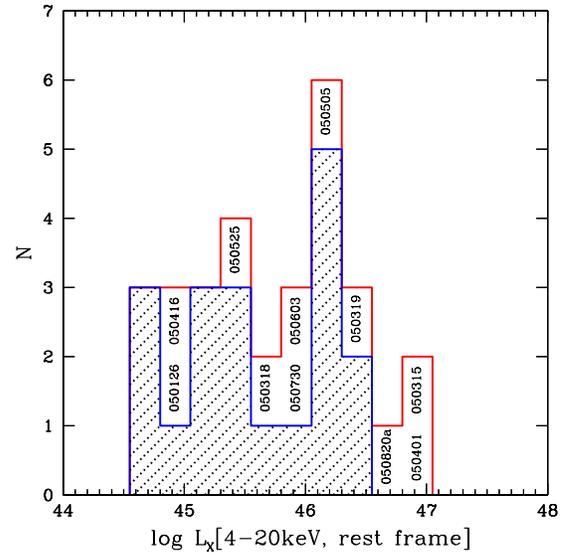,angle=0,width=8.8cm}
\vskip -0.5 true cm
\caption{In this histogram we added (to the histogram shown in Fig.
  \ref{isto_x}), the X--ray [4--20 keV] luminosities 
  12 hours (rest frame) after the trigger of the 11 GRBs       
  discovered by SWIFT whose intrinsic luminosities could be calculated
  from the data found in Nousek et al. (2005; for GRB 050730, see
  Perri et al. 2005; 
  for GRB 050820, see Page et al. 2005).
} 
\label{istox_swift}
\end{figure}

\section{SWIFT bursts}

At the time of writing (November 2005), there are 18 long GRBs
detected by SWIFT for which the redshift has been spectroscopically
determined. 
We list them in Table \ref{swift}, together with their redshifts, 
galactic extinction, and, when possible, the 
calculated optical luminosities (at 12 hours rest frame).
We also list the X--ray luminosities calculated in the
(rest frame) 4--20 keV band, 12 hours after trigger.

We alert the reader that for all these GRBs but GRB 050401, GRB 050525
and GRB 050730
there are no information yet about 
the optical extinction in the rest frame of the source.
Therefore the listed value of $L(\nu_{\rm R}) $ is not
corrected for extinction in the rest frame, and is k--corrected assuming
$\beta=1$ (except for GRB 050401, for wich we used $A_V^{\rm
  host}=0.67$ and $\beta=0.5$ (Watson et al., 2005) GRB 050525, for which we used 
the values given in Blustin et al. 2005. 
For GRB 050730 Starling et al. (2005) give a measured value of $\beta=1$).
The optical light curves the 14 SWIFT bursts with redshift 
are shown in Fig. \ref{luce_swift},
superposed to the light curves (green or light grey symbols) of the other bursts.
For the remaining 4 SWIFT bursts with known redshift we could not find
enough information 
in the literature for plotting their light curve.
In Fig. \ref{isto_swift} we  show the histogram of the optical
luminosities after the addition of the 14 SWIFT bursts for which
we could calculate this quantity. 
As can be seen, although the behavior of the light curve of
some of the SWIFT bursts is peculiar, their $L(\nu_{\rm R})$
at 12 hours is entirely consistent with the  distribution
found previously.

The X--ray luminosity distribution with the addition of these SWIFT
bursts is shown in Fig. \ref{istox_swift}.
Note that the SWIFT bursts appear, on average, more luminous in X--rays
than the pre--SWIFT bursts.
This is likely due to the fact that the average redshift of SWIFT
bursts is somewhat larger than the average redshift of the other bursts.
($\langle z \rangle =2.3\pm 1.3$ for the bursts listed in Tab. \ref{swift}
vs $\langle z \rangle =1.5\pm 0.9$ for the bursts listed in Tab. 4.)

We conclude that the indications coming from the first SWIFT
bursts with known redshift are strongly confirming the picture
presented in this paper. Despite the difference in average redshift,
and despite the broadening of the X--ray luminosity
distribution, the clustering of the large majority of the optical 
luminosities is confirmed.
In addition, the SWIFT bursts also confirm the existence of a dichotomy
of the optical luminosity distribution, with the presence
of an underluminous family.

\section{Discussion}

The main results of our study is the finding of a clustering 
of the optical luminosities of the afterglows of GRBs.
That is, bursts with widely different isotropic gamma--ray emitted
energies are nevertheless similar in their optical output.

This result is unexpected for several reasons:
i) The optical luminosities, for the time of interests,
do not dominate the bolometric radiated output;
ii) Contrary to the X--ray frequencies, likely to be 
above the cooling frequency $\nu_c$, the optical frequencies
are likely to be smaller than $\nu_c$.
This is confirmed by the simultaneous SED shown in Fig. \ref{sed1}
and Fig. \ref{sed2}. 
This implies that the optical emission does depend on the density of the 
interstellar medium, $n$.
A range in the values of $n$ should then contribute to increase
the dispersion of the optical luminosities.
iii) Similarly to the X--ray luminosities, also $L_{\rm opt}$ 
depends from the product of $E_{\rm k, iso}$ and a function
of the equipartition parameters $\epsilon_e$ and $\epsilon_B$.
The observed clustering implies a corresponding ``clustering"
of the values of the isotropic kinetic energy and of the
equipartition parameters.

\subsection{The ``standard" external shock synchrotron model}

In order to understand what observed, we should 
investigate the implications of these two facts:
i) for the majority of bursts $\nu_c$ is between the optical
and the X--ray band after a few hours (to a day) from trigger;
ii) the distribution of the optical luminosities  is narrower
than the distribution of X--ray luminosities.
We here very briefly discuss these facts in the framework
of the standard external shock synchrotron model

As previously noted by Panaitescu \& Kumar  (2000, 2001, 2002)
having $\nu_c$ between the optical and X--ray bands
a day after the trigger implies a relatively small value
of $\epsilon_B$ (and $n$).
For convenience, we report here Eq. 27 (for homogeneous ISM density)
and Eq. 28 (for a $r^{-2}$ wind profile)
of Panaitescu \& Kumar (2000) for $\nu_c$:
\begin{equation}
\nu_c \, =\,
3.7\times 10^{14} E_{53}^{-1/2} n^{-1} (Y+1)^{-2} 
\epsilon_{B,-2}^{-3/2} t_d^{-1/2}
\quad {\rm Hz}
\end{equation}
\begin{equation}
\nu_c \, =\,
3.4\times 10^{14} E_{53}^{1/2} A_{*}^{-2}\, (Y+1)^{-2} \,
\epsilon_{B,-2}^{-3/2} \, t_d^{1/2} \,\,
\quad {\rm Hz}
\end{equation}
where the notation $Q=10^x Q_x$ is adopted.
$E$ is the isotropic kinetic energy of the fireball,
$t_d$ is the time after trigger measured in days and
$Y$ is the Comptonization parameter.
For the wind case it is assumed that $n(r)=Ar^{-2}$ and
$A_*$ is the value of $A$ when setting 
$\dot M =10^{-5} M_\odot$ yr$^{-1}$
and a wind velocity $v=10^3$ km s$^{-1}$.
From the above equations, values of $\nu_c$ close to 
$10^{16}$ Hz require $\epsilon_B \sim 10^{-3}$ or less.
The possible dependence of $\nu_c$ from the slope of the electron 
energy distribution is hidden in the $(Y+1)$ term.
This term is important if $\epsilon_B$ is below some critical
value (see discussion in Panaitescu \& Kumar 2000).

\begin{figure}
\vskip -0.5 true cm
\psfig{figure=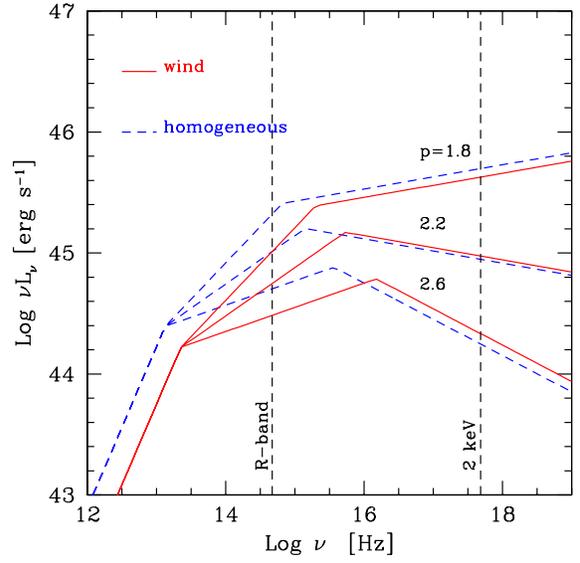,angle=0,width=8.8cm}
\vskip -0.5 true cm
\caption{
Examples of spectra calculated using the prescriptions of 
Panaitescu \& Kumar 2000, at 12 hours after trigger.
Dashed lines corresponds to a homogeneous ISM case 
(with density $n=1$ cm$^{-3}$); solid lines to a wind profile
of the density (with $\dot M=3\times 10^{-6} M_\odot$ yr$^{-1}$
and wind velocity $v=10^3$ km s$^{-1}$. 
The models differ for the assumed values of $p$ (as labelled).
}
\label{spectrum2}
\end{figure}

In order to find the simplest possible reason for
the clustering of the optical luminosities, we used
again the analytical prescriptions of Panaitescu \& Kumar (2000)
to construct light curves and spectra at a given time.
Fig. \ref{spectrum2} shows some examples of spectra
calculated at 12 hours after trigger, assuming for all cases
the same kinetic energy ($E=10^{53}$ erg), the same 
$\epsilon_B=10^{-3}$ value, the same $\epsilon_e=10^{-1}$,
and external density ($n=1$ cm$^{-3}$ for the homogeneous ISM case
and $\dot M=3\times 10^{-6} M_\odot$ yr$^{-1}$ and $v=10^3$ km s$^{-1}$
for the wind case).
What changes is only the slope of the electron distribution $p$.
As can be seen we indeed obtain in this case that the optical luminosities
are distributed in a much narrower range than the X--ray luminosities.
This is due to the fact that the cooling frequency changes 
when changing $p$ as a result of Compton losses being important,
decreasing for smaller $p$. 
Note also that this is true both for the homogeneous and the wind case.

We stress that this example is only illustrative, and it does not
pretend to give an exhaustive explanation of our results, since there
can be other solutions in which more than one parameter is changing.
Keeping this in mind, the observed clustering of the optical luminosities
would then require that the kinetic (isotropically equivalent)
energy is distributed in a narrow range, as are the equipartition
parameters.
Furthermore, $\epsilon_B$ (and/or the density $n$) should be
small, and the Compton $Y$ parameter relatively large.

\subsection{Dark Bursts}

Our results can help to understand why a significant fraction
of bursts with detected X--ray afterglows are not detected
in the optical, even now in the SWIFT era, which allows a very fast
reaction and optical observations both onboard through UVOT 
and on ground through several robotic telescopes.
Although our samples is still limited, it appears that there
is a family of optically underluminous objects (dimmer by an order 
of magnitude with respect to the average luminosity of the main family).
This bursts are the obvious candidates to be missed in the optical,
especially in the presence of absorption in the host galaxy
(Nardini et al. in prep.).

We can wonder if these GRBs appear underluminous because of
an underestimation of the intrinsic absorption.
In this respect, we note that for GRB 011121 three different authors
are all estimating a null value of $A_V^{\rm host}$, while they differ
somewhat for the value of the galactic extinction.
To be conservative, we have taken the largest value of those.
For GRB 021211, there are two estimates of $A_V^{\rm host}$, and 
we used the conservative choice, taking the largest value.
For GRB 980613 there is only one estimate of $A_V^{\rm host}$,
but a constraint on the maximum possible value of $A_V^{\rm host}$
comes from its SED. 
From Fig. \ref{sed1} it can be seen that the optical luminosity
of this bursts cannot be larger than a factor of ten from what
is plotted, if we require a smooth joining of the extrapolated
optical/X--ray spectra.
This would barely bring this burst to the faint end of the luminous family.
However, in this case the SED would appear anomalous, because the
peak of the spectrum would be in the optical--near--UV (contrary
to the majority of the other bursts), and the optical spectral index
resulting from such a large correction would be
flatter than $\beta_o=0.5$, again in contrast with the other bursts.

We do not know yet if these few underluminous bursts are the
tip of the iceberg of a much more numerous populations, and we do not
know the corresponding spread in luminosities.
We believe that SWIFT will clarify this point.

\section{Conclusions}

The main results of our study are:

\begin{itemize}

\item The optical luminosities of GRB afterglows, calculated at the same rest frame
time, show an unexpected tight clustering, with most (21/24) of the optical
luminosities spanning less than one order of magnitude around a mean value
of $\log L_{\nu_R}=30.65$.

\item A minority (3/24) of GRBs form a separate dimmer family, with an optical
luminosity one order of magnitude less than the one of the more numerous
family.

\item
These results have been obtained considering all bursts with known
redshift and optical extinction in the host galaxy,
but the inclusion of the recently detected SWIFT bursts
(of still unknown extinction in the host) is fully consistent
with these findings, and reinforces them.

\item The optical luminosity distribution appears narrower than the
X--ray luminosity distribution of the same bursts, calculated at the same
rest frame time.

\item The isotropic optical luminosities are not correlated with $E_{\rm \gamma, iso}$.

\item The X--ray isotropic luminosity correlates (even if not strongly)
with the isotropic prompt emitted energy $E_{\rm \gamma, iso}$, but this
is simply due to the dependence of both quantities on the aperture angle
of the jet.  The collimation corrected prompt energy and X--ray luminosity
are not correlated.

\item The optical to X--ray SEDs of our bursts show that for most
of the objects the entire observed emission is due to the same
(synchrotron) process (after several hours to a day after trigger, rest frame time).
In a $\nu L_\nu$ plot, the peak frequency lies in between the optical and
the X--ray bands. The peak frequency can be identified with the
cooling frequency.

\item Our results are quite unexpected, and their interpretation
is not obvious. One possibility points towards the importance of 
changing the slope of the electron energy distribution, while
the other parameters are more constant.


\end{itemize}

\begin{acknowledgements}
We thank the Italian MIUR for founding (Cofin grant
2003020775\_002).
\end{acknowledgements} 

\section{Appendix}

References for the data plotted in Fig. \ref{f_obs} and Fig. \ref{lc1}.
\begin{itemize}
\item
GRB 970508: Garcia et al, 1998; Sokolov et al.,
 1998; Vietri et al., 1998.  
\item
GRB 971214: Diercks et al., 1998.
\item
GRB 980613: Hjorth  et al., 2002.  
\item
GRB 980703: Bloom  et al.,
 1998, Castro--Tirado et al., 1999, Vreeswijk et al., 1999.
\item
GRB 990123: Odewahn et al., 1999, (IAUC 7094); Zhu et al, 1999,
 (IAUC 7095); Zhu et al., 1999, (GCN 204); Lachaume  et al., 1999,
 (IAUC 7096); Ofek et al., 1999, (GCN 210); Maury et al.,1999,
 (IAUC 7099); Garnavich  et al., 1999, (GCN 215); Masetti et al.,
 1999, (GCN 233); Sagar et al., 1999, (GCN 227); Yadigaroglu  et
 al., 1999, (GCN 242); Veillet, 1999, (GCN 253); Veillet, 1999, (GCN260). 
\item
GRB 990510: Harrison et al., 1999; Israel et al., 1999.   
\item
GRB 991216: Garnavich et al., 2000; Halphern et al., 2000.  
\item
GRB 000301c: Jensen et al., 2001; Bhargavi et al., 2000. 
\item
GRB 000418: Berger et al., 2001.  
\item
GRB 000911: Price et al., 2002; 
 Lazzati et al., 2001; Masetti et al., 2005.
\item
GRB 000926: Fynbo  et al., 2001; Price et al., 2001.  
\item
GRB 010222: Galama et al., 2003.  
\item
GRB 010921: Price  et al., 2002, ApJ, 571,L121.  
\item
GRB 011121: Greiner et al., 2003; Garnavich et al., 2003.   
\item
GRB 011211: Jakobsson et al., 2003.  
\item
GRB 020124: Hjorth et al., 2003; Berger et al., 2002.  
\item
GRB 020405: Price et al., 2002, (GCN 1326); Price et al. 2002
 (GCN 1333); Gal--Yam et al., 2002, (GCN 1335); Hjorth, 2002,
 (GCN 1336).
\item
GRB 020813: Laursen \& Stanek, 2003; Urata et al., 2003;
 Li et al., 2003.  
\item
GRB 021004: Bersier et al., 2003; Pandey et al., 2003; Holland  et
 al., 2003.   
\item
GRB 021211: Holland et al., 2004; Pandey et al.,
 2003; Li et al., 2003, b; Fox et al., 2003.  
\item 
GRB 030323: Vreeswijk, et al., 2004.  
\item 
GRB 030226: KLose S. et al., 2004; Pandey et al., 2004.
\item 
GRB 030329: Lipkin et al., 2004; Torii et al., 2003; 
 Torii, 2003, (GCN 1986); Rykoff, 2003, (GCN 1995); Gal--Yam,
 2003, (GCN 1999), Klose et al., 2003, (GCN 2000); Burenin et al.,
 2003, (GCN 2001); Lipunov et al., 2003, (GCN 2002);
 Martini et al., 2003, (GCN 2012); Masi et al., 2003, (GCN 2016);
 Halpern et al., 2003, (GCN 2021); Zharikov  et al., 2003,
 (GCN 2022); Burenin et al., 2003, (GCN 2024); Rumyantsev et al,
 2003, (GCN 2028); Klose et al., 2003, (GCN 2029);
 Bartolini et al., 2003, (GCN 2030); Lipkin et al., 2003,
 (GCN 2034); Stanek et al., 2003, (GCN 2041); Lipkin et al.,
 2003, (GCN 2045); Burenin et al., 2003, (GCN 2046); Zeh et
 al., 2003, (GCN 2048); Lipkin et al., 2003, (GCN 2049);
 Burenin et al., 2003, (GCN 2051);  Burenin et al., 2003,
 (GCN 2054); Fitzgerald et al., 2003, (GCN 2056); Price, 2003,
 (GCN 2058); Lipkin et al., 2003, (GCN 2060); Li et al., 2003,
 (GCN 2063); Pavlenko et al., 2003, (GCN 2067); Fitzgerald et
 al., 2003, (GCN 2070); Price et al., 2003 (GCN 2071); Cantiello  et
 al., 2003, (GCN 2074); Zharikov et al., 2003, (GCN 2075);
 Ibrahimov et al., 2003, (GCN 2077);  Burenin et al., 2003,
 (GCN 2079); Sato et al., 2003, (GCN 2080); Pavlenko et al., 2003,
 (GCN 2083); Ibrahimov et al., 2003, (GCN 2084); Khamitov et
 al., 2003, (GCN 2094); Lee et al., 2003, (GCN 2096);  Pavlenko et al.,
 2003, (GCN 2097); Ibrahimov et al., 2003, (GCN 2098); 
 Urata et al., 2003, (GCN 2106); Khamitov et al., 2003,
 (GCN 2108); Lyuty et al., 2003, (GCN 2113); Suzuki et al., 2003,
 (GCN 2116); Khamitov et al., 2003, (GCN 2119); Rumyantsev et al.,
 2003, (GCN 2146); Ibrahimov et al., 2003, (GCN 2160); Zharikov et al.,
 2003, (GCN 2171); Semkov, 2003, (GCN 2179);  Ibrahimov et al., 2003,
 (GCN 2191); Kindt, et al., (GCN 2193); Khamitov et al., 2003,
 (GCN 2198); Ibrahimov et al., 2003, (GCN 2219); Pizzichini at al.,
 2003, (GCN 2228); Stanek et al., 2003 (GCN 2244); Stanek et al., 2003,
 (GCN 2259);  Burenin et al., 2003, (GCN 2260); Zharikov et al., 2003,
 (GCN 2265); Ibrahimov et al., 2003, (GCN 2288);  Khamitov et
 al., 2003, (GCN 2299).
\item 
GRB 030429; Jakobsson et al., 2004. 
\end{itemize}

\end{document}